\documentclass[12pt,preprint]{aastex}

\shorttitle{The Spitzer Young Stellar Cluster Survey}
\shortauthors{Megeath et al.}

\begin{document}


\title{Initial Results from The Spitzer Young Stellar Cluster Survey}


\author{S. T. Megeath\altaffilmark{1}, L. E. Allen\altaffilmark{1},
  R. A.  Gutermuth\altaffilmark{2},J. L. Pipher\altaffilmark{2}, P. C. Myers\altaffilmark{1}, 
N. Calvet\altaffilmark{1}, L. Hartmann\altaffilmark{1}, J. Muzerolle\altaffilmark{3} \& G. G. Fazio\altaffilmark{1}}

\altaffiltext{1}{Harvard-Smithsonian Center for Astrophysics, Mail Stop 42, 60 Garden Street, Cambridge, MA 02138 (tmegeath@cfa.harvard.edu)}
\altaffiltext{2}{Department of Physics and Astronomy, University of Rochester, Rochester, NY 14627}
\altaffiltext{3}{Steward Observatory, University of Arizona, 933 N. Cherry Ave. Tucson, AZ 85721}

\begin{abstract}
We report initial results from IRAC observations of four young stellar
clusters. These regions are part of a larger {\it Spitzer} survey of 31
young stellar groups and clusters within 1 kpc of the Sun.  In each of
the four clusters, there are between 39 and 85 objects with colors
inconsistent with reddened stellar photospheres.  We identify these
objects as young stars with significant emission from circumstellar
dust. Applying an analysis developed in a companion paper
\citep{allen04}, we classify these objects as either pre-main sequence
stars with disks (class~II) or protostellar objects (class~I).  These
show that the sites of recent star formation are distributed over
multi-parsec size scales.  In two clusters, Cepheus C and S140, we
find protostars embedded in filamentary dark clouds seen against
diffuse emission in the IRAC bands.

\end{abstract}

\keywords{stars:pre-main sequence --- stars: formation ---
planetary systems:protoplanetary disks ---- infrared:stars ---
ISM:clouds --- ISM:individual(\objectname{Cepheus C}, \objectname{S171}, 
\objectname{S140}, \objectname{NGC~7129})}

\section{Introduction}

In the past 30 years, infrared observations have revolutionized our
understanding of star formation.  IRAS and ISO observations of young
stellar objects showed that young stellar objects with disks and
infalling protostellar envelopes exhibit distinctive infrared spectral
energy distributions (SEDs) in the mid and far-IR \citep{w&l83, als87,
  osorio03}.  Ground-based surveys of molecular clouds using near-IR
detector arrays found that young stars typically form in clusters
\citep{lada92,carp00}.  The {\it Spitzer} Space Telescope promises to
further revolutionize the study of star formation by providing the
capability to image young stellar groups and clusters in the mid-IR
with the sensitivity to detect young stars down to the hydrogen
burning limit and below.  With this new capability, we can build upon
the legacies of IRAS, ISO and numerous ground-based near-IR
observations by probing the SEDs of stars, brown dwarfs and protostars
in young stellar clusters out to distances of 1 kpc or greater.

As part of the Guaranteed Time Observations of the IRAC instrument
team, we will image 31 young stellar groups and clusters with the MIPS
and IRAC instruments.  These have been selected from a catalog of 63
star forming regions within 1 kpc of the Sun containing 10 or more members
\citep{porr03}.  In parallel, we are mapping 7 sq. degrees in the
Orion molecular clouds.  These surveys will sample the full continuum
of multiple star forming regions in the nearest kiloparsec, from small
groups of stars in Taurus to the rich Orion Nebula and Mon R2
clusters.

This paper reports on initial results for four young stellar clusters
in our sample with properties between those of Taurus and Orion: S140,
S171, Cepheus C and NGC7129.  These regions are at similar distances,
but span a range of FIR luminosities, molecular gas masses, and
cluster membership (Table~\ref{properties}).  In each cluster, we use
the IRAC photometry to identify young stars with disks and protostars.
We base our classification on the results of a companion paper by
\citet{allen04}, in which the IRAC colors of the observed young stars
are compared to colors derived from models of stars with disks
or infalling envelopes.

\begin{table}
\begin{center}
\caption{Young Stellar Cluster Properties\label{properties}}
\begin{tabular}{llllll}
\tableline\tableline
Source & Distance$^1$  & IRAS$^1$ & Molecular$^{2}$ &  Cluster & Cluster \\
 & (pc)  & Luminosity & Cloud Mass& Radius$^3$ & Membership$^4$ \\
 &  & (L$_{\odot}$) & (M$_{\odot}$) & (pc) & (stars) \\
\tableline
Cepheus C & 700 & 106   & 1200  & 0.3$^5$  & 42$^5$ \\
S171      & 850 & 60    & 870   & 0.33$^6$ & 28$^6$ \\
S140      & 900 & 20560 & 1290  & 0.24$^7$ & 34$^7$ \\
NGC 7129  & 1000 & 1360 & 980   & 0.51$^8$ & 80$^8$ \\
\tableline
\tablenotetext{1}{\citet{ridg03}}
\tablenotetext{2}{Masses 
from $^{13}$CO ($1 \rightarrow 0$)  data
in \citet{ridg03}}
\tablenotetext{3}{Cluster radius derived from an analysis of star counts vs.
radius from the cluster center  \citet{gute04}}
\tablenotetext{4}{Number of detected stars within cluster radius minus
the expected number of background stars.  The background star density
is estimated from a nearby control field}
\tablenotetext{5}{Determined from $K'$ mosaic from Hodapp (1994)}
\tablenotetext{6}{Number of sources down to $K = 15.5$}
\tablenotetext{7}{Number of sources down to $K = 14.5$}
\tablenotetext{8}{Number of sources down to $Ks = 16$; \citet{gute04}}
\end{tabular}
\end{center}
\end{table}

\section{Observations and Data Analysis}

Observations of Cepheus C, S171, NGC7129 and S140 were obtained on
2003 December 19, 23, and 24, respectively, with the InfraRed Array
Camera (IRAC) \citep{fazio04}.  Each region was mapped in a $4 \times
3$ grid resulting in a $10' \times 15'$ field with coverage in all
four IRAC wavelength bands.  The 12 second high dynamic range mode was
used to obtain two frames at each position, one with an exposure time
of 0.4 seconds and one with an exposure time of 10.4 seconds.  The map
was repeated four times with small offsets, resulting in a total
integration of 41.6 seconds per pixel after co-adding the longer
duration frames.  The observations were processed with the SAO IRAC
Pipeline (SIP) and mosaics were created with a custom IDL program.

Source finding and photometry was performed using Gutermuth's Photvis
1.08 which incorporates Landsman's IDLPHOT package into an interactive
GUI \citep{land93}. To subtract out a spatially varying nebulosity, we
used an aperture radius of 2 pixels ($2.4''$) and a sky annulus
extending from 2 pixels ($2.4''$) to 6 pixels ($7.2''$).  The
calibration of the data was performed with in-flight observations of
IRAC standard stars, and aperture corrections were determined for each
band from observations of an IRAC standard star. In this paper, we
consider only sources detected in all four bands.  The standard
deviations returned by IDLPHOT includes a contribution of $\sigma (\pi
r^2)^\frac{1}{2}$, which is added in quadrature, where $r$ is the aperture
radius in pixels and $\sigma$ is the standard deviation of the pixel
values in the sky annulus.  Consequently, bright, well detected stars
can have large uncertainties if they are coincident with a bright,
spatially varying nebulosity.  Since regions of active star formation
often have bright nebulosity in the 5.8 and 8~$\mu$m bands, we adopt a
maximum photometric uncertainty of 0.25~mag for the 5.8 and 8.0~$\mu$m
data, as compared to 0.2~mag for the 3.6 and 4.5~$\mu$m data.  We
include only stars brighter than 14~mag at 5.8~$\mu$m and 13~mag at
8.0~$\mu$m; at these magnitudes the typical uncertainties are 0.2~mag
or less.  The resulting median uncertainties are 0.01, 0.01, 0.04 and
0.06~mag, and only 3\% of the stars in our sample have uncertainties
greater than 0.2~mag in the 5.8~$\mu$m and/or 8.0~$\mu$m bands.  Each
image was visually inspected, and nebulous knots misidentified as
stars and close double stars were rejected.

\section{Classifying young stars using the IRAC color plane}

In an analysis of IRAS photometry of young stellar objects in nearby
dark clouds, \citet{w&l83} classified young stellar objects as class
I, II and III by the slope of their SEDs.  The classification scheme
was placed in an evolutionary context by \citet{als87}, who modeled the
SEDs as stars surrounded by dusty disks and envelopes. By comparing
these models to the observed SEDs, they identified class~I objects as
protostars with infalling envelopes and class~II objects as stars with
disks.  Class~III objects have the SEDs of stellar photospheres.  The
three classes are thought to form an evolutionary sequence, with young
stars evolving from class~I to class~II and finally to class~III
objects \citep{kenyon95}.

\begin{figure}
\epsscale{0.75}
\plotone{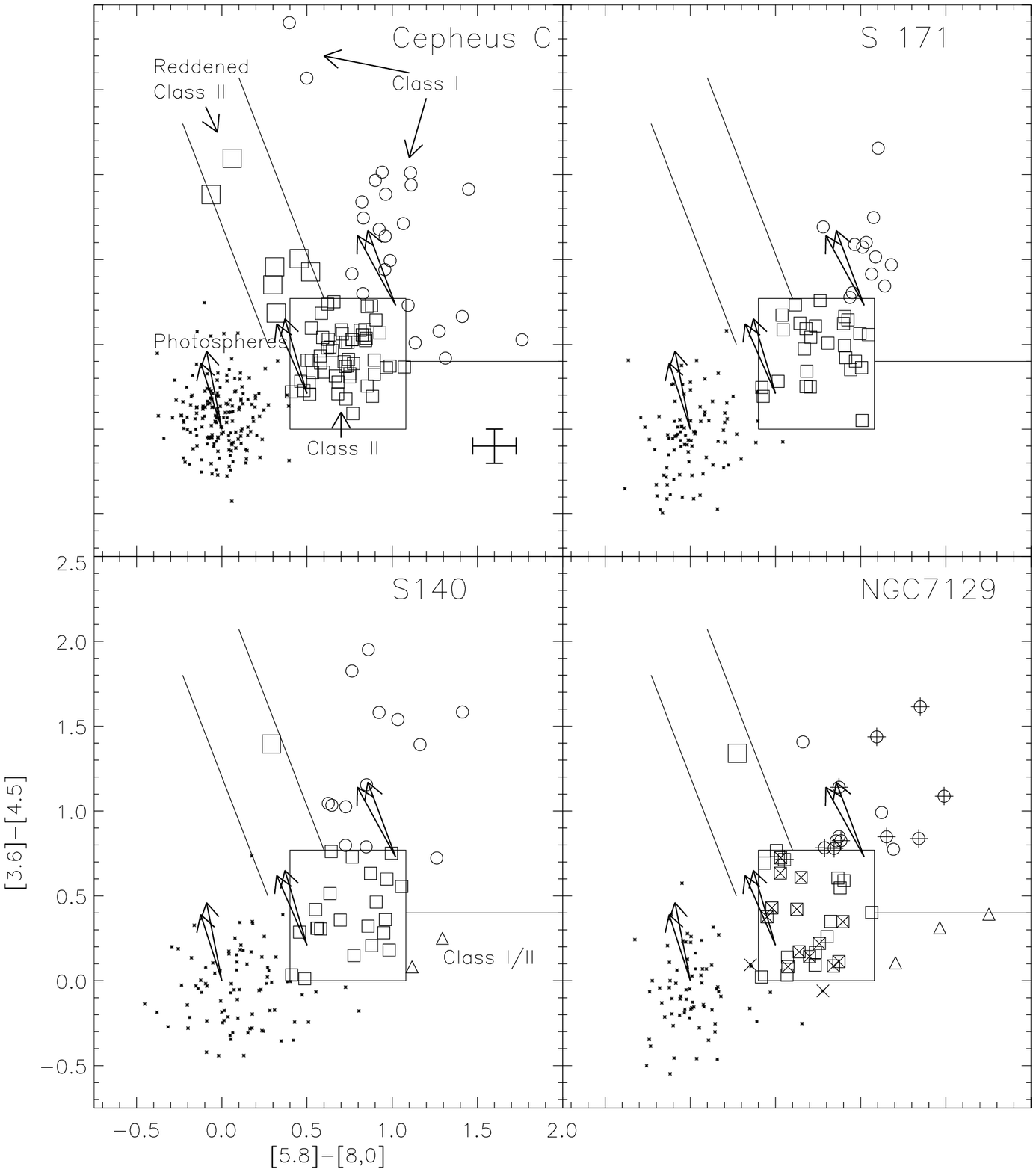}
\vskip 0.5 in
\caption{The IRAC color-color diagram is displayed for all four
  clusters. Using a selection criteria outlined in the text and in
  \citet{allen04}, the squares are identified as class~II sources, the
  large square are reddened class~I sources, and the circles are class
  I sources.  The two parallel lines border the positions of sources
  we identify as reddened class~II objects.  In two of the regions,
  S140 and NGC 7129, we mark the class~I/II sources (which share the
  characteristics of both class~I and class~II sources) with
  triangles; the horizontal line above the triangles shows the adopted
  division between class~and classI/II sources.  We show reddening
  vectors for $A_V = 30$ derived from the \citet{draine84} and
  \citet{mathis90} extinction laws, in each case the \citet{draine84}
  vector points to the left of the \citet{mathis90} vector. The
  vectors were calculated for Vega (at position 0,0), a young star
  with disks taken from the models of \citet{paola04} (0.2,0.5), and a
  flat spectrum source (0.73,1).  Twenty-four sources in NGC7129 were
  classified using IRAC photometry combined with MIPS 24~$\mu$m
  photometry by \citet{muze04}; to display these classifications we
  overplot a plus sign for Class~I objects and an $X$ for class~II
  objects.  The error bars in the Cepheus C plot show the median
  uncertainty in the colors for all the sources in all four clusters.
  A conservative 0.1~mag calibration uncertainty was added in
  quadrature to the median photometric uncertainty. 
\label{color}}

\end{figure}

Instead of performing an analysis directly on the infrared SEDs, we
use the colors derived from the four IRAC bands to classify each young
star.  Figure~\ref{color} shows that the $[3.6]-[4.5]$ and
$[5.8]-[8.0]$ colors extend over 2~magnitudes.  We display two
reddening vectors derived from the extinction law of \citet{mathis90}
and from the optical constants of \citet{draine84}.  We have applied
these two extinction laws to three different sources: a model of Vega
\citep{cohen92}, a star with disks taken from an ensemble of models by
\citet{paola04}, and an idealized flat spectrum source with $\lambda
F_{\lambda} = constant$.  Since the IRAC 8~$\mu$m band overlaps the
silicate feature, reddened photospheres would appear increasingly blue
in the $[5.8]-[8.0]$ color; the slope of the reddening vector depends
on the extinction law, the intrinsic spectrum of the source, and the
amount of extinction.

Centered at $([3.6]-[4.5],[5.8]-[8.0])=(0,0)$ are the sources with the
colors of stellar photospheres.  This region of the color plane may
include foreground and background stars as well as diskless (class
III) pre-main sequence stars.  The distribution of sources appears
elongated in the vertical direction; this is due to a combination of
reddening and the blue $[3.6]-[4.5]$ colors of background giants with
strong CO absorption in the 4.5~$\mu$m band.  In all four clusters
there is a concentration of sources centered near
$([3.6]-[4.5],[5.8]-[8.0])=(0.7,0.5)$, which we outline with a box in
Figure~\ref{color}.  These colors can be reproduced by models of disks
around young, low-mass stars \citep{allen04,whit03}.  The range in
colors can be explained largely by variations in inclination and
accretion rate \citep{allen04}.  Based on the range of colors exhibited
by the models, we classify objects with $0.0 > ([3.6]-[4.5]) > 0.8$
and $0.4 > ([5.8]-[8.0]) > 1.1$ as class~II objects.  Although the
models also predict stars with $([5.8]-[8.0]) < 0.4$, we have adopted
this limit to reliably distinguish class~II sources from ``colorless''
class~III/foreground/background stars.  Given that only one source in
all four regions with $([5.8]-[8.0]) < -0.4$, and assuming that the
photometric scatter is symmetric with respect to $([5.8]-[8.0]) = 0$,
we estimate the percentage of colorless stars misidentified as Class
II sources is less than 1\%.

Sources with $([3.6]-[4.5]) > 0.8$ and/or $([5.8]-[8.0]) > 1.1$ cannot
be explained by reddened class~II objects.  These objects have colors
similar to those derived from models of protostellar objects with
infalling dusty envelopes \citep{allen04, whit03}.  We identify these
sources as class~I objects (we do not distinguish between class 0 and
class 1 objects in our current analysis).  Several sources in each
cluster exhibit colors which are not consistent with models of class~I
objects, class~II objects or reddened photospheres.  In
Fig.~\ref{color}, several sources show higher $[3.6]-[4.5]$ colors
than class~II objects, but lower $[5.8]-[8.0]$ colors than class~I or
class~II objects.  Based on the predicted slope of the reddening
vector, we identify these nine sources as reddened class II objects.
Five sources (three in NGC 7129 and two in S140) have $([5.8]-[8.0]) >
1.1$, consistent with class~I objects, but $([3.6]-[4.5]) < 0.4$,
which is lower than that predicted by class~I models
\citep{allen04}. Because these sources share the properties of class~I
and II sources, we refer to these as class~I/II sources.  These
sources are in regions with bright, structured 8~$\mu$m nebulosity,
and these may be class~II sources in which compact knots and filaments
of nebulosity are contributing to the signal in the 8~$\mu$m band.

There are several factors that may lead to the incorrect
classification of sources.  First, models show some overlap between
class~I sources and class~II sources in the IRAC color plane
\citep{allen04}. We classify the sources in this overlap region as
class~II. Second, photometric scatter may alter the classification of
a few sources.  Third, some of the sources which we have identified as
class~I objects may be in fact reddened class~II objects.  Finally,
unresolved binaries can result in the incorrect classification of
sources. The Herbig Be star LkH$\alpha$234 in NGC 7129 is wrongly
identified as a class~I object in the IRAC color plane; this
misclassification is due to a protostellar companion which is detected
in ground-based mid-IR observations but is unresolved by IRAC
\citep{cabrit99}.  Background planetary nebulae, AGB stars, and
galaxies may also be misidentified as young stars or protostars
\citep{whit03}; however, the spatial distribution of sources discussed
in the following section indicates that the amount of background
contamination is small.

By combining MIPS 24~$\mu$m photometry with the IRAC photometry of NGC
7129, \citet{muze04} identify class~I and class~II objects in NGC 7129
by using the slope of the $1-24$~$\mu$m SEDs.  This data provides an
independent check of our classification using the IRAC color plane.
For the twenty four sources common to both samples (Muzerolle et
al. required detection in {\it three} IRAC bands {\it and} the MIPS
24~$\mu$m band), the classification by \citet{muze04} is almost
entirely consistent with ours (see NGC7129 plot in Fig~\ref{color}).
The one discrepant source has colors consistent with both 
class~II objects and low luminosity class~I objects \citep{allen04}.

The number of class~I and II sources in each cluster is given in
Table~\ref{yso}.  Our census of class~I and class~II objects has been
demonstrated to be incomplete in NGC7129 by \citet{gute04}. By
requiring detection in only the IRAC $4.5$~$\mu$m band and combining
the IRAC photometry with ground-based $J$ and $H$-band photometry, they
find a total of 84 objects with circumstellar disks in NGC7129 within
a $9'\times9'$ field.

\begin{table}
\begin{center}
\caption{Statistics on Source Detection and YSO Classification\label{yso}}
\begin{tabular}{lllllll}
\tableline\tableline
Source & Total & Class I & Class II & Reddened & Class I/II & Total \\
& Sources &  &  & Class II &  & YSOs \\
\tableline 
Cepheus C & 263 & 23 & 55 & 7 & 0 & 85 \\
S171      & 134 & 12 & 27 & 0 & 0 & 39 \\
S140      & 122 & 13 & 22 & 1 & 2 & 38 \\
NGC 7129  & 119 & 14 & 27 & 1 & 3 & 45 \\
\tableline
\end{tabular}
\end{center}
\end{table}

\section{The Distribution of Young Stars}

We show IRAC mosaics of all four clusters in Figures~\ref{imagea} and
\ref{imageb}.  In preparation for the {\it Spitzer} young stellar
cluster survey, \cite{ridg03} mapped each of these regions in the
$^{13}$CO and C$^{18}$O ($J = 1 \rightarrow 0$) transition.  We overlay
contours of the the C$^{18}$O emission which is an excellent tracer of
the structure of the molecular gas in star forming regions
\citep{gold97}.  In comparison, the emission from the more abundant
$^{13}$CO molecule is stronger and more extended, but has a higher
optical depth and is less sensitive to the detailed structure in the
gas \citep{ridg03}.  We find 72\% of the class I and 56\% of the class
II sources fall within the detected C$^{18}$O emission, and 93\% of
the class~I and 86\% of the class~II sources fall within the detected
$^{13}$CO emission. These percentages are qualitatively consistent
with the class~I sources being younger than the class~II sources.

The Cepheus C cluster was first identified in a near-IR survey by
\citet{hoda94}.  In addition to the near-IR cluster, which is located
toward the center of Figure~\ref{imagea}, the IRAC data show class~I
and II sources distributed over a 3 pc diameter region.  The molecular
gas traced by the C$^{18}$O is visible in the IRAC images as
filamentary dark clouds obscuring a diffuse nebulosity extending
across the entire mosaic. Two class~I objects appear outside the
C$^{18}$O emission; $^{13}$CO emission is found toward both of these
sources.

The S171 cluster is a compact cluster of young stars found in a
bright-rimmed cloud.  The OB stars of the Cepheus IV association,
which illuminate the bright rim, are outside the image and to the
south.  The funnel shaped surface of the cloud emits strongly in the
IRAC 5.8 and 8.0~$\mu$m bands and is clearly seen in
Figure~\ref{imagea}; the $^{13}$CO map shows the molecular gas
extending to the edge of the funnel.  S171 contains a cluster of young
stars near the edge of the cloud; with a dense group of 5 class~I
sources at northern apex of the cluster.  This morphology suggests
that star formation is being triggered by a photoevaporation driven
shock-wave propagating into the cloud, as first proposed for this
region by \citet{sug95}.  In addition to the stars in the cluster,
there are 6 class~II and 2 class~I objects spread throughout the
molecular cloud.  The presence of these stars suggests that a
distributed mode of star formation is also occurring in the cloud.

S140 is also a bright-rimmed cloud and the one region of massive star
formation in our sample; as in case of S171, the illuminated surface
is clearly delineated in Figure~\ref{imageb}.  A group of at least
three early B stars are thought to have recently formed in this region
\citep{evans89, prei02}; these three sources are saturated in our data
and appear as a bright extended source near the center of the image.
The bright emission from the central stars and the extended nebulosity
make the detection of sources in all four bands difficult.  We
identify three class~I and three class~II objects within a 0.5~pc
radius of the young massive stars; a near-IR star counts analysis
reveals 34 members within this radius
(Table~\ref{properties}). Extending east into the molecular clouds is
a filamentary dark cloud seen in absorption against the diffuse
emission; in these filaments are five class~I sources, and three
class~II sources.  Near the eastern edge of the map are two additional
class~I objects which appear outside the contours of the C$^{18}$O
map; however, $^{13}$CO emission is detected toward these sources
\citet{ridg03}. To the southwest of the cluster and outside the bright
rimmed cloud are 9 class~II objects, a class~I object, and a
class~I/II object.  These may be young stars which emerged into the
HII region when their natal molecular gas was overtaken by the
advancing ionization front.

Near-IR observations show that NGC~7129 contains a cluster of over 80
young stars.  Many of these stars are coincident with a bright
reflection nebula and cannot be detected in all four IRAC bands; hence
the dense clustering of sources found in the near-IR is not apparent
in Figure~\ref{imageb}.  The molecular cloud, as delineated by the
C$^{18}$O contours, wraps around the reflection nebula \citep{ridg03};
the distribution of Class~I and II sources in Figure~\ref{imageb}
shows that star formation is ongoing in this cloud.  There are eight
class~I objects outside the contours of the C$^{18}$O emission.
Emission in $^{13}$CO is detected toward all but three of these
sources: the northernmost class~I object is outside the $^{13}$CO map
and no molecular gas is detected toward the two class~I objects west
of the reflection nebula.  \citet{gute04} and \citet{muze04} discuss
IRAC and MIPS observations of NGC~7129 in more detail.

\section{Discussion}

In all four clusters, the young stars and protostars identified by
their excess emission in the mid-IR are distributed over multi-parsec
distances.  In contrast, the diameters of the clusters identified by
near-IR star counts are typically 1 pc or less \citep{lada03}. This
suggests that a significant fraction of stars in each star forming
region form outside the dense clustered regions identified in star
counts analyses. \citet{gute04} find that half of the stars in
NGC 7129 are located in a halo outside the cluster core.  Furthermore,
star formation in NGC~7129 is continuing in the halo, while the
molecular gas has been dispersed toward the cluster core.

The distribution of sources in each region is strikingly different.
In Cepheus C, the structure of the molecular cloud breaks up into
distinct mid-IR dark cores, and several distinct concentrations of
stars are also apparent to the eye.  The observed distribution of gas
and stars in this region is similar to the hierarchical morphologies
generated in numerical models of star formation in turbulent clouds
\citep{bonnel03}.  IRAC images of NGC~7129 show a dense cluster of
primarily class~II sources surrounded by a more extended halo of class
I and II objects.  Both S171 and S140 contain compact clusters at the
edges of bright-rimmed clouds.  These varied mophologies hint that
environmental factors, such as the presence of external OB stars, may
play a significant role in the formation of clusters.

\acknowledgments

This work is based on observations made with the {\it Spitzer} Space
Telescope, which is operated by the Jet Propulsion Laboratory,
California Institute of Technology under NASA contract 1407. Support
for this work was provided by NASA through Contract Number 1256790
issued by JPL/Caltech.  Support for the IRAC instrument was provided
by NASA through Contract Number 960541 issued by JPL.

\clearpage

\clearpage

\begin{figure}
\epsscale{.9}
\caption{Images of Cepheus C (top) and S171 (bottom) constructed from
  the IRAC 3.6 (blue), 4.5~$\mu$m (green) and 8.0~$\mu$m (red) images.
  The contours are the maps of C$^{18}$O ($1 \rightarrow 0$) emission
  from \citet{ridg03}.  The C$^{18}$O observations have an angular
  resolution of $50''$. We show the position of each young star
  identified in the IRAC color-color diagram. We mark class~II sources
  with squares, reddened class~II objects with large squares, and
  class~I objects with circles.  The class~I/II sources are marked by
  triangles.
\label{imagea}} 
\end{figure}

\begin{figure}
\epsscale{.9}
\caption{Images of S140 (top) and NGC7129 (bottom)
  using the the same scheme described in Figure~2
\label{imageb}} 
\end{figure}

\end{document}